\documentclass[11pt,twoside]{article}

\usepackage{asp2014}

\aspSuppressVolSlug
\resetcounters

\begin{document}

\title{Distributed Real-Time Data Stream Analysis for CTA}

\author{Kai Br\"ugge$^1$, Alexey Egorov$^1$, Christian Bockermann$^1$, Katharina Morik$^1$ and Wolfgang Rhode$^1$}
\affil{$^1$TU Dortmund, Dortmund, Germany; \newline \email{kai.bruegge@tu-dortmund.de} \newline \email{alexey.egorov@tu-dortmund.de}}

\paperauthor{Kai~Br\"ugge}{kai.bruegge@tu-dortmund.de}{}{TU Dortmund}{Astroparticle Physics}{Dortmund}{NRW}{44227}{Germany}
\paperauthor{Alexey~Egorov}{alexey.egorov@tu-dortmund.de}{}{TU Dortmund}{Artificial Intelligence Group}{Dortmund}{NRW}{44227}{Germany}
\paperauthor{Christian~Bockermann}{christian.bockermann@tu-dortmund.de}{}{TU Dortmund}{Artificial Intelligence Group}{Dortmund}{NRW}{44227}{Germany}
\paperauthor{Katharina~Morik}{katharina.morik@tu-dortmund.de}{}{TU Dortmund}{Artificial Intelligence Group}{Dortmund}{NRW}{44227}{Germany}
\paperauthor{Wolfgang~Rhode}{wolfgang.rhode@tu-dortmund.de}{}{TU Dortmund}{Astroparticle Physics}{Dortmund}{NRW}{44227}{Germany}

\begin{abstract}

  Once completed, the  Cherenkov Telescope Array (CTA) will be able to map the gamma-ray sky in a wide energy range from several tens of GeV to some hundreds of TeV and will be more sensitive than previous experiments by an order of magnitude.
  It opens up the opportunity to observe transient phenomena like gamma-ray bursts (GRBs) and flaring active galactic nuclei (AGN).  In order to successfully trigger multi-wavelength observations of transients, CTA has to be able to alert other observatories as quickly as possible.  Multi-wavelength observations are essential for gaining insights into the processes occurring within these sources of such high energy radiation.

  CTA will consist of approximately 100 telescopes of different sizes and designs.
  Images are streamed from all the telescopes into a central computing facility on site.
  During observation CTA will produce a stream of up to 20 000 images per second. Noise suppression and feature extraction algorithms are applied to each image in the stream as well as previously trained machine learning models.
  Restricted computing power of a single machine and the limits of network's data transfer rates become a bottleneck for stream processing systems in a traditional single-machine setting.
  We explore several different distributed streaming technologies from the Apache Big-Data eco-system like Spark, Flink, Storm to handle the large amount of data coming from the telescopes.
  To share a single code base while executing on different streaming engines we employ  abstraction layers such as the streams-framework.
  These use  a high level language to build up processing pipelines that can transformed into the native pipelines of the different platforms.
  Here we present results of our investigation and show a first prototype capable of analyzing CTA data in real-time.

\end{abstract}

\section{The Cherenkov Telecope Array}
Once completed, the  Cherenkov Telescope Array (CTA)  will be able to map the gamma-ray sky
in a wide energy range from several tens of GeV to some hundreds of TeV and will be more sensitive
than previous experiments by an order of magnitude.
CTA will consist of approximately 100 imaging air cherenkov telescopes (IACTs) of three different sizes.
Telescope data will be streamed via network from the telescopes to a central computing facility on-site.
Cherenkov telescopes record light produced by particle showers induced in the upper atmosphere by cosmic rays.
While the cosmic ray flux is approximately isotropic over the sky, gamma rays of cosmic origin can be pinpointed back to its source.
Filtering air showers produced by cosmic rays while keeping those produced by gamma rays is the big challenge in IACT data analysis.

\section{Real Time Analysis}

One of CTA's main goals is monitoring the sky for transient events.
These include Gamma-Ray Burst (GRBs) events and Active Galactic Nuclei (AGNs).
To gain more understanding about the physics working in GRBs and AGN it is vital to perform multi-wavelength observations.
In case a GRB is detected,  CTA can alert other experiments to trigger observations in other wavelength bands.
The data aquisition system will supply the real time analysis (RTA) with calibrated images of each triggered telescope.
Features for classication/regression are calculated on each image in the triggered events.
Simulated and labeled data is used to train the models using the Python machine learning library \texttt{scikit-learn}~\citep{sklearn}.
These models are then used for filtering of comsic ray showers and estimation of primary particle energy.
The trained \texttt{scikit-learn} models are converted into the PMML~\citep{pmml} format using the \texttt{sklearn2pmml}~\citep{sklearn2pmml} library.
This way the stored model can be shared between programming languages and applied to the data stream from the telescopes.

The expected event rate of CTA will be between $10000$ and $20000$ events per second depending on deployment
and pointing direction~\citep{trigger}. Every real time processing system for CTA data will have to handle that data rate.

\articlefigure{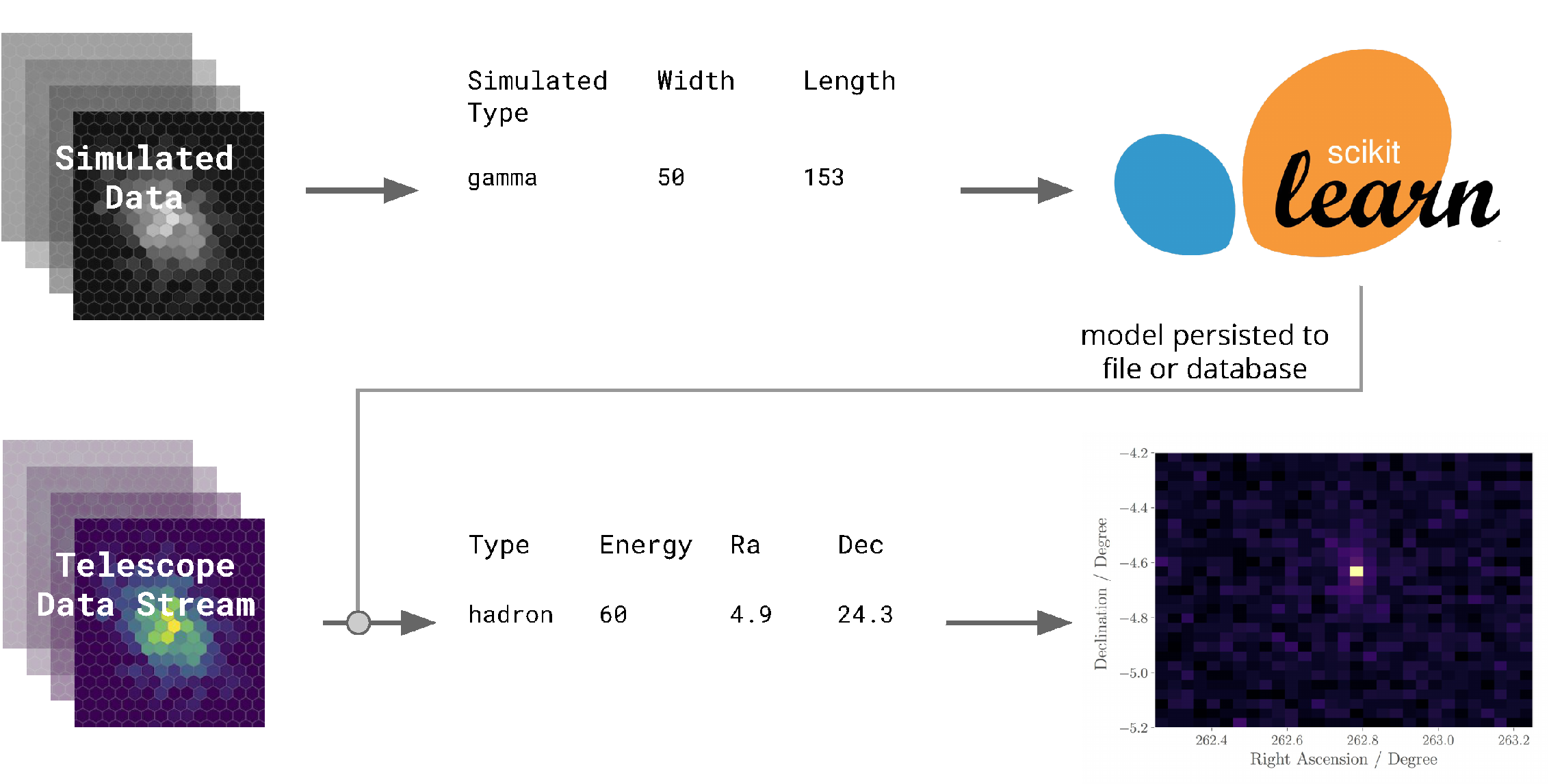}{fig:overview}{Overview of the analysis process. We use \texttt{scikit-learn} to train
models on simulated data and apply these models to the data stream coming from the telescopes to get results in near real time.}

\section{Machine Learning Performance}

Supervised machine learning is used to separate signal from background and to estimate the energy of a recorded event.
Features for classification/regression are calculated from raw input images.
We trained a Random-Forest~\citep{rf} classifier with 200 trees on simulated data to separate signal events from background events.

The trained model is then applied to each image from the telescopes in the data stream.
Figure~\ref{fig:machine-learning-perf} presents the results of the model application.
The plot on the left shows the performance of background suppression for the three different telescope types.
The predictions for each telescope are then averaged to get a combined prediction for the entire event.
This improves background rejection significantly as shown in the right image.

Energy estimation is performed in a similar manner. A Random Regression Forest is trained on the same simulated data.
The resulting model is then applied to the data stream and predictions are averaged for all images in a single event.

\articlefiguretwo{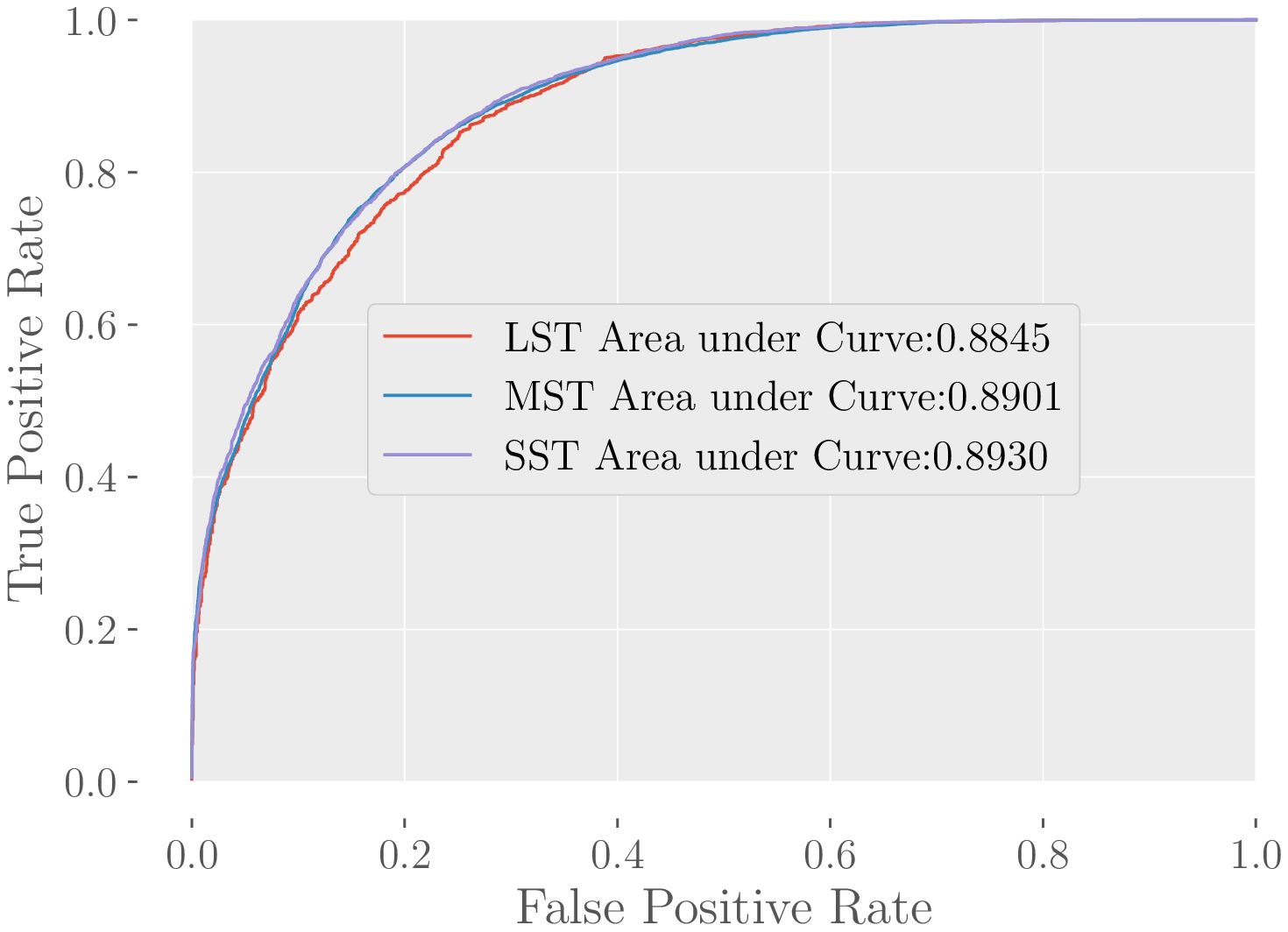}{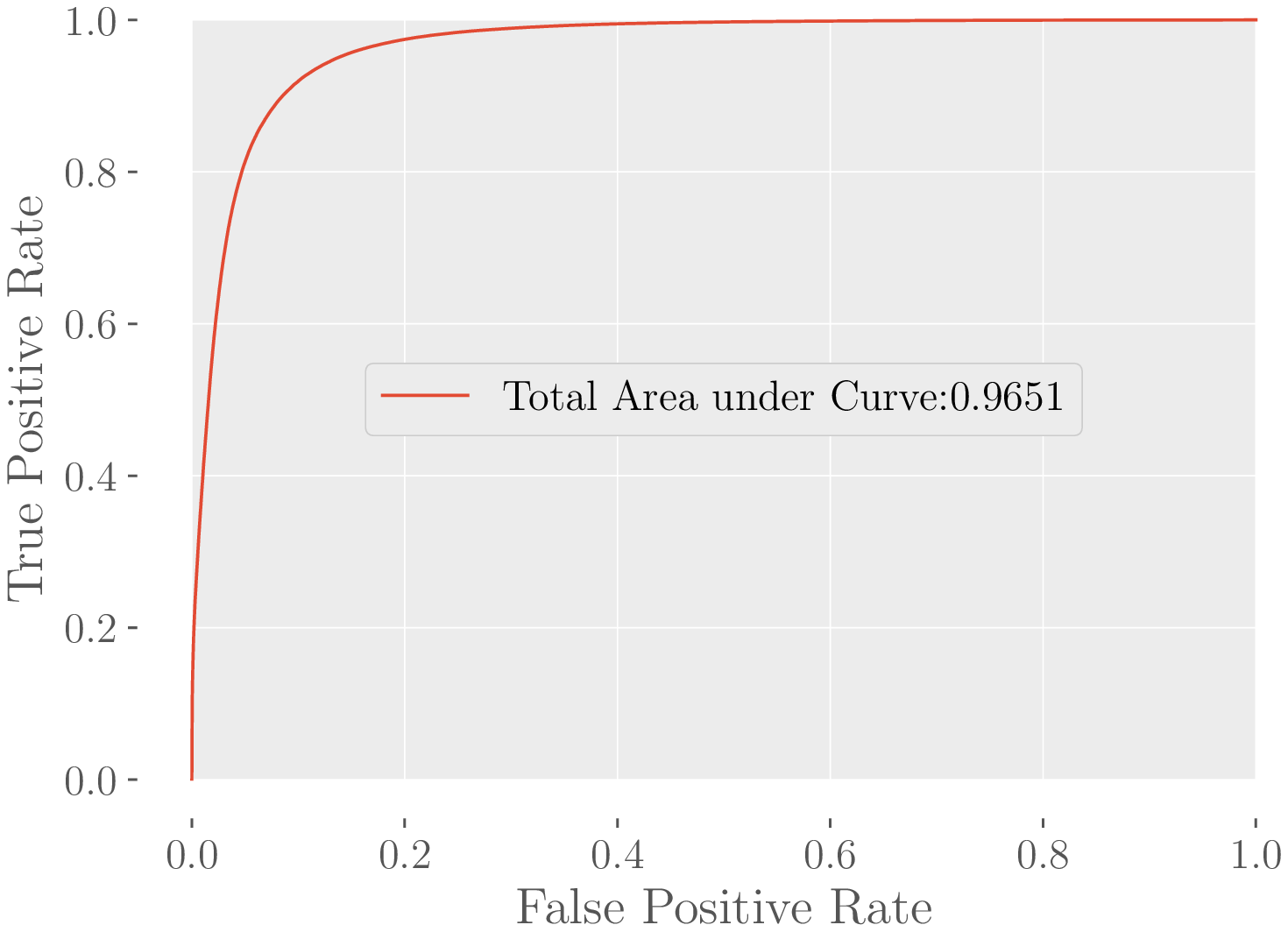}{fig:machine-learning-perf}{\emph{Left:} Performance of background suppression for the
different telescope types. The small size telescope (SST), the mid size telescope (MST) and the large size telescope (LST).
\emph{Right:} Improved background suppression of a combined predictor when averaging the predictions for all telescopes in
a CTA event.}

\section{Runtime Performance}

Frameworks for distributed stream processing such as Apache Storm \citep{storm} or Apache Flink \citep{flink}
allow for workload distribution with fault tolerance and high availability mechanisms to recover from hardware or network failures.
Tuning Apache Spark for fast streaming applications is more complicated due to its micro-batch architecture.
We compare the runtime of Storm and Flink on a single core to measure the impact of the overhead these frameworks produce.
To share a single code base while executing on different streaming engines we employ the \texttt{streams-framework}~\citep{streams}
as an abstraction layer.
This way the analysis pipeline is only defined once, but can be easily executed on top of different distributed streaming engines.
The left image in Figure~\ref{fig:dist-frameworks} shows that Flink shows less runtime overhead compared to Storm.

We continue our experiments on Flink due to faster processing, easier setup and a more comfortable high-level API compared to Storm.
Any real-time processing system for CTA has to be able to handle the data rate of up to $20000$ events per second and provide analysis results
within 30 seconds according to the official CTA requirements.

The right Figure~\ref{fig:dist-frameworks} presents the evaluation of a full CTA pipeline executed on top of Flink.
For this test a machine with $24$ physical CPU cores was used.
The datarate goes up to approximately $14000$ events per second on this single machine.
We use a simple self hosted Flink cluster on two machines with 24 cores each to scale the process up.
This way event rates of more than $20 000$ events per second are achieved.

\articlefiguretwo{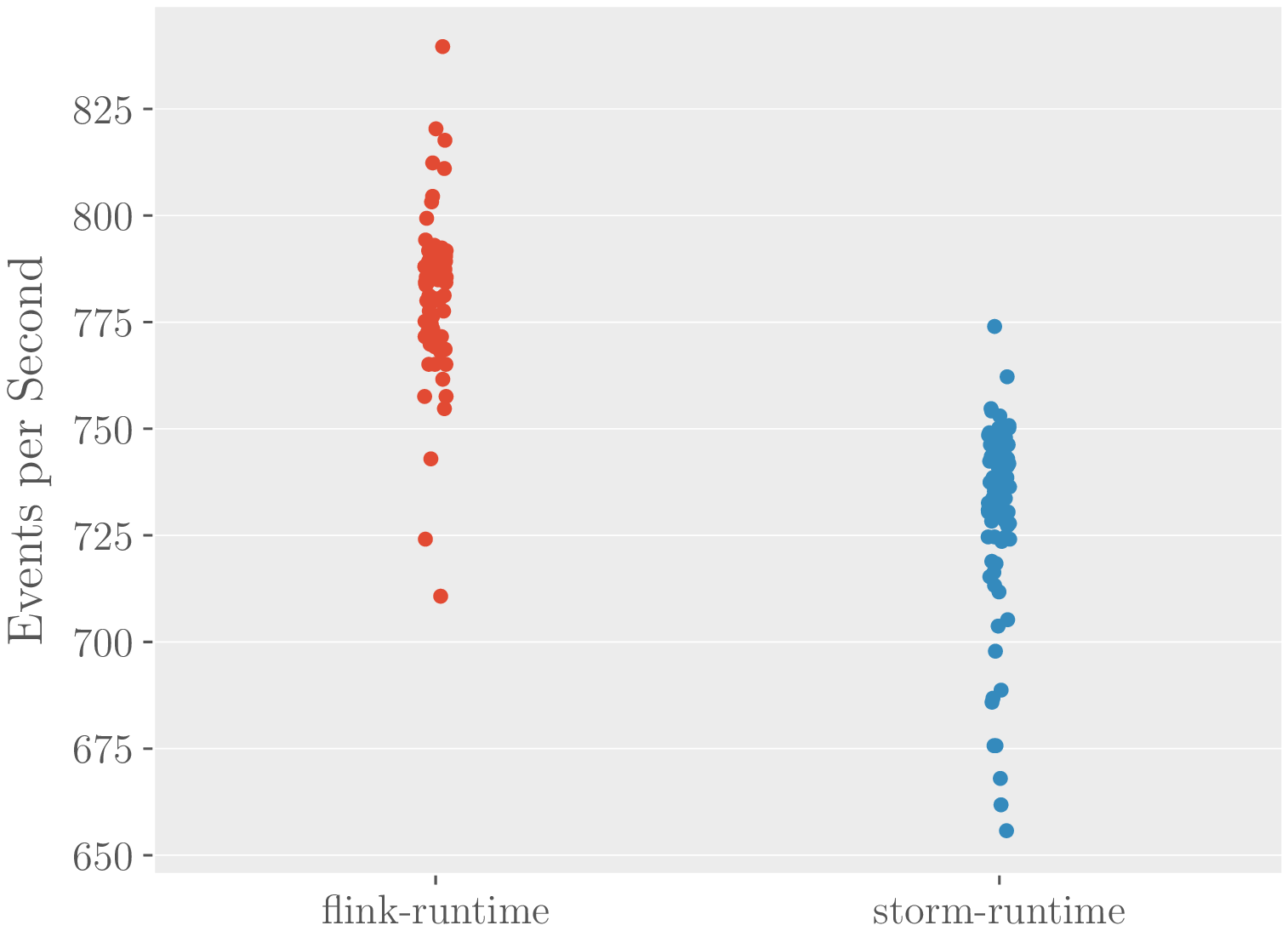}{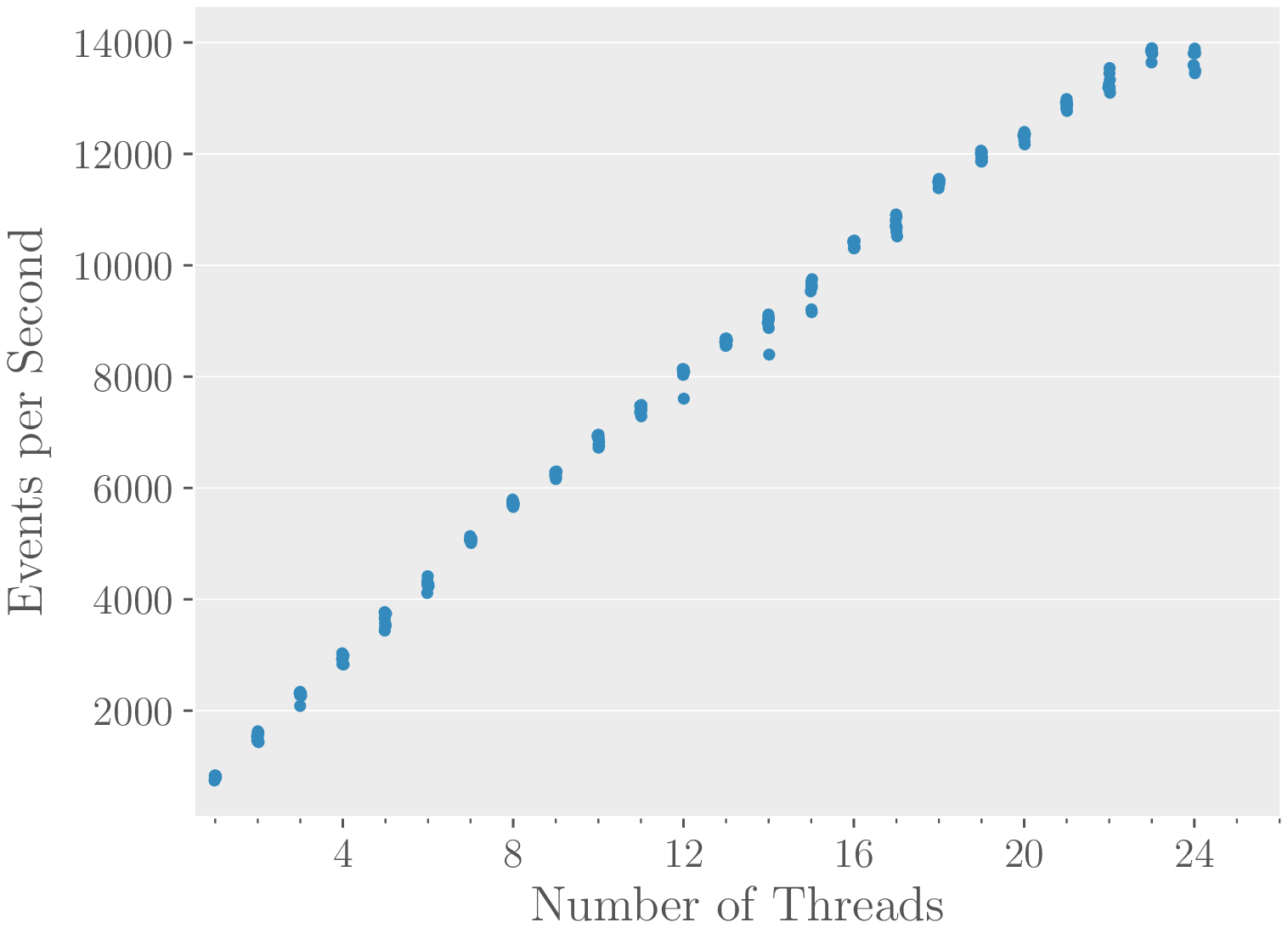}{fig:dist-frameworks}{\emph{Left:} Overhead produced by distributed processing frameworks.
\emph{Right:} Throughput of CTA analysis pipeline executed on multiple cores.}

\acknowledgements Part of this work is supported by Deutsche Forschungsgemeinschaft (DFG) within the Collaborative Research Center SFB 876 "Providing Information by Resource-Constrained Analysis", project C3.

\bibliography{P6-63}

\end{document}